\documentclass{aipproc}
\layoutstyle{8x11double}
\usepackage{mathptm}
\begin{document}%%%%%%%%%%%%%%%%%%%%%%%%%%%%%%%%%%%%%%
\title{
Multifractal Analysis for the Dynamical Heterogeneity 
in Strongly Correlated Many-Body Systems
}

\author{O. Narikiyo and W. Sakikawa}{
  address={Department of Physics, 
           Kyushu University, 
           Fukuoka 810-8560, Japan}
}

\begin{abstract}
By calculating the non-equilibrium parameter 
of the probability distribution function and 
the singularity spectrum of multifractal 
we have quantified the dynamical heterogeneity 
in strongly correlated many-body systems. 
\end{abstract}

\maketitle%%%%%%%%%%%%%%%%%%%%%%%%%%%%%%%%%%%%%%%%%%%%

\section{Introduction}%%%%%%%%%%%%%%%%%%%%%%%%%%%%%%%%
\vspace{-1mm}

We have numerically studied the dynamical heterogeneity 
in strongly correlated many-body systems, 
for example, the critical spin state\cite{SN1}, 
the supercooled liquid near the glass transition\cite{SN2} 
and the turbulence\cite{SN3}. 
These systems are scale-invariant and multifractal. 
In this paper we briefly review 
how we quantify the dynamical heterogeneity 
and discuss in detail the turbulence simulation data 
recently obtained. 

In order to quantify the dynamical heterogeneity 
we have two measures. 
One is the non-equilibrium parameter $q$ 
of the R\'enyi-Tsallis distribution function. 
The other is the singularity spectrum $f(\alpha)$. 
These two measures are closely related 
and important ingredients of the multifractal analysis. 

In the numerical study of the critical spin state\cite{SN1} 
we have found that the $q$-parameter represents 
the degree of non-equilibrium. 
We have actually quantified the deviation of the q-parameter 
from the equilibrium value according to the spatio-temporal 
scale of the observation. 

In the numerical study of 
the the supercooled liquid near the glass transition\cite{SN2} 
we have found that the broken-bond distribution 
which reflects the nature of 
the cooperatively rearranging region 
is well described by the singularity spectrum $f(\alpha)$ 
of multifractal. 
The width of the spectrum becomes broader 
as approaching the glass transition. 
Such a broadening is similar to that observed in the numerical study 
of the Anderson-localization transition. 

\vspace{-1mm}
\section{Turbulence\ \ Simulation}%%%%%%%%%%%%%%%%%%%%%%%%%
\vspace{-1mm}

In the numerical study of the turbulence\cite{SN3} 
we have adopted the lattice Boltzmann method. 
Our numerical turbulence on $200^3$ cubic lattice points 
is sustained by a random forcing 
and the Reynolds number is about 500. 

%------------------------------------------------------------
\begin{figure}[htbp]
\includegraphics[width=70mm, height=70mm]{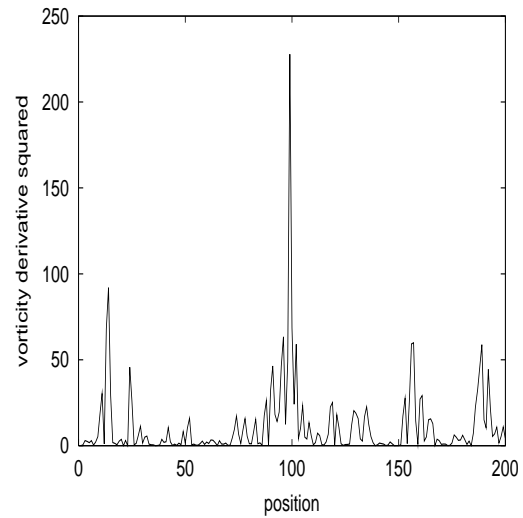}
\caption
{
The intermittent distribution 
of the strength of the vortices.
}
\label{myid:fig:1}
\end{figure}
%------------------------------------------------------------

\vspace{-1mm}
\subsection{mean-field\ analysis}%%%%%%%%%%%%%%%%%%%%%%%%%%%
\vspace{-1mm}

Turbulence is one of the typical phenomena 
with multiscale motions. 
Each phenomena of a scale strongly couples with 
all the other scales of turbulent motion. 
In order to analyze such a system 
a scale-dependent entropy, 
the so-called $\epsilon$-entropy $h(\epsilon)$, 
works well. 
For example, the time series of the velocity field in turbulence 
leads to a non-trivial scaling relation, 
$ h(\epsilon) \propto \epsilon^{-3} $, 
expected from Kolmogorov's scaling. 
Here $\epsilon$ is the scale of the observation. 
The existing experimental data are consistent with this scaling. 
By our numerical experiment simulating the Navier-Stokes equation 
we have shown the consistency of the scaling. 
In contrast to the evaluation of the energy spectrum, 
which is usually employed for testing Kolmogorov's scaling 
and determined by the two-point correlation function, 
Kolmogorov's scaling is easily observed in the $\epsilon$-entropy, 
since it is determined by the mean of the exit time 
from the observation window 
and fluctuations are averaged out. 
Such a unifractal scaling is a mean-field description and 
fluctuations can be taken into account in a multifractal analysis 
as shown in the following. 
In the above mentioned mean-field description 
the velocity difference in temporal and spatial directions 
have the same fractal scaling exponent. 
Thus we have confirmed 
Kolmogorov's scaling and Taylor's hypothesis at the same time. 

\vspace{-1mm}
\subsection{fluctuation\ analysis}%%%%%%%%%%%%%%%%%%%%%%%%%%%
\vspace{-1mm}

In order to analyze the fluctuations 
we use two measures, 
the non-equilibrium parameter $q$ 
of the R\'enyi-Tsallis distribution function 
and the singularity spectrum $f(\alpha)$. 

The probability distribution function (PDF) for 
the velocity or vorticity field is non-Gaussian. 
The non-Gaussianity is quantified by the $q$-parameter. 
The strong non-Gaussianity is observed 
at small spatio-temporal scale 
comparable to that for the coherent vortex 
in the PDF for the velocity or vorticity difference 
between two space-time points. 
By using the wavelet denoising 
we have clarified that the non-Gaussianity or 
dynamical heterogeneity 
results from the existence of the coherent vortex 
which is a strongly-correlated non-equilibrium region. 

The strength of the vortices is shown in Fig. 1 
to be intermittent. 
The spatial distribution of the coherent vortices 
is quantified by the singularity spectrum of multifractal 
$f(\alpha)$ as shown in Fig. 2. 

%------------------------------------------------------------
\begin{figure}[htbp]
\includegraphics[width=70mm, height=70mm]{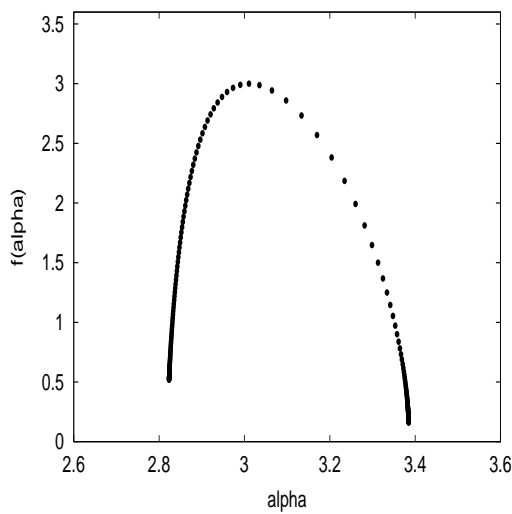}
\caption
{
The singularity spectrum $f(\alpha)$ of the spatial distribution 
of the coherent vortices.
}
\label{myid:fig:2}
\end{figure}
%------------------------------------------------------------

The remainder of this subsection is devoted to some speculations. 
As seen in our previous study\cite{SN2} 
the width of the singularity spectrum 
depends on the degree of intermittency 
so that we expect some Reynolds-number dependence of the spectrum. 
It has been discussed by many authors 
and should be claryfied in future systematic study. 
In our present study 
the density of the vortex is dilute. 
Thus the singularity spectrum in Fig. 2 
describes the spatial distribution of relatively free vortices. 
As the Reynolds number is increased the density increases. 
In this case 
the interaction among vortices becomes important and  
the correlation length of the vorticity fluctuation becomes large. 
While we can observe the vortex 
only as an individual elementary excitation 
in our numerical experiment, 
some collective excitation is expected to dominate 
at higher Reynolds number. 
In the limit of divergently large Reynolds number 
the correlation length becomes divergently large 
so that we can expect full scale-invariance. 
We can find a resemblance to the case of the scaling theory 
in polymers where an ideal scaling relation is realized for dense solutions 
where polymers are strongly entangled. 
In the limit of high Reynolds number 
each boxes counting the coherent vortex 
for calculating $f(\alpha)$ 
will be filled by almost equal number of vortices 
so that intermittency will disappear and 
unifractal Kolmogorov's scaling will prevail. 

\vspace{-1mm}
\subsection{summary\ for\ turbulence\ simulation}%%%%%%%%%%%%%%%%%%%%
\vspace{-1mm}

Numerical simulation data, in real space and time, 
for a forced turbulence 
on the basis of the lattice Boltzmann method 
have been analyzed by unifractal and multifractal schemes. 

Our new findings are summarized into two points. 
First in the unifractal analysis using the exit-time statistics 
we have verified Kolmogorov's scaling and Taylor's hypothesis 
at the same time. 
Second in the analysis 
using the R\'enyi-Tsallis PDF and the wavelet denoising 
we have clarified that the coherent vortices sustain 
the power-law velocity correlation in the non-equilibrium state. 

Finally in the multifractal analysis 
it is clarified 
that the intermittent distribution 
of the coherent vortices in space-time 
is described as a multifractal. 

%%%%%%%%%%%%%%%%%%%%%%%%%%%%%%%%%
\end{document}